\documentclass[aps,prb,twocolumn,superscriptaddress,amsmath,amssymb]{revtex4}

\usepackage{amssymb}
\usepackage{bm}
\usepackage{epsfig}
\usepackage[ansinew]{inputenc}
\usepackage{graphicx}

\begin{document}

\title{Unveiling charge density wave quantum phase transitions by x-ray diffraction}

\author{F. B. Carneiro}
\affiliation{Instituto de F\'{\i}sica, Universidade do Estado do Rio de Janeiro, 20550-900, Rio de Janeiro, RJ, Brazil}
\affiliation{Centro Brasileiro de Pesquisas F\'{\i}sicas, 22290-180, Rio de Janeiro, RJ, Brazil}
\author{L. S. I. Veiga}
\affiliation{Deutsches Elektronen-Synchrotron (DESY), Notkestra$\beta$e 85,  22607, Hamburg, Germany}
\affiliation{London Centre for Nanotechnology and Department of Physics and Astronomy, University College London, Gower Street, WC1E 6BT, London, United Kingdom}
\author{J. R. L. Mardegan}
\affiliation{Deutsches Elektronen-Synchrotron (DESY), Notkestra$\beta$e 85,  22607, Hamburg, Germany}
\author{R. Khan}
\affiliation{Centro Brasileiro de Pesquisas F\'{\i}sicas, 22290-180, Rio de Janeiro, RJ, Brazil}
\author{C. Macchiutti}
\affiliation{Centro Brasileiro de Pesquisas F\'{\i}sicas, 22290-180, Rio de Janeiro, RJ, Brazil}
\author{A. L\'{o}pez}
\affiliation{Instituto de F\'{\i}sica, Universidade do Estado do Rio de Janeiro, 20550-900, Rio de Janeiro, RJ, Brazil}
\author{E. M. Bittar}
\email{bittar@cbpf.br}
\affiliation{Centro Brasileiro de Pesquisas F\'{\i}sicas, 22290-180, Rio de Janeiro, RJ, Brazil}

\begin{abstract}
We investigate the thermal-driven charge density wave (CDW) transition of two cubic superconducting intermetallic systems Lu(Pt$_{1-x}$Pd$_{x}$)$_2$In and (Sr$_{1-x}$Ca$_x$)$_3$Ir$_4$Sn$_{13}$ by means of x-ray diffraction technique. A detailed analysis of the CDW modulation superlattice peaks as function of temperature is performed for both systems as the CDW transition temperature $T_{\rm CDW}$ is suppressed to zero by an non-thermal control parameter. Our results indicate an interesting crossover of the classical thermal-driven CDW order parameter critical exponent from a three-dimensional universality class to a mean-field tendency, as $T_{\rm CDW}$ vanishes. Such behavior might be associated with presence of quantum fluctuations which influences the classical second-order phase transition, strongly suggesting the presence of a quantum critical point (QCP) at $T_{\rm CDW}=0$. This also provides experimental evidence that the effective dimensionality exceeds its upper critical dimension due to a quantum phase transition.
\end{abstract}

\maketitle

\section{Introduction}

Charge-density-wave (CDW) and its relation to superconductivity and quantum criticality have been a subject of intense research in the recent years. In cuprates high-temperature superconductors, the interplay between the CDW and superconducting states is a matter of intense debate and still an open question \cite{Cu1,Cu2,Cu3,Cu4}. Intermetallic compounds displaying a CDW phase transition also shows a variety of other interesting physical phenomena, such as superconductivity, which can be enhanced as the charge instability is tuned to zero by pressure and/or chemical doping \cite{qcp3413,qcp34132,CDWQCP,2H-NbSe,Lu}. The suppression of the CDW, which sets in below $T_{\rm CDW}$, by a non-thermal control parameter, might be related with the enhancement/emergence of the superconducting state, associated with quantum fluctuations near a quantum critical point (QCP) in the limit $T_{\rm CDW}\rightarrow0$. For several CDW systems, however, the evidence that quantum fluctuation does in fact play an important role in the suppression of the CDW are not clear. The observation that the superconducting temperature transition $T_c$ is favoured as $T_{\rm CDW}$ decreases may be solely related to the closing of the partially gapped CDW, which increases the density of states at the Fermi energy $N(E_f)$ benefiting the superconducting state as $T_c\propto e^{-1/[N(E_f)V_{eff}]}$, where $V_{eff}$ is an effective attractive interaction \cite{CDWQCP}.

Classical thermal-driven continuous or second-order phase transitions are well known to be defined by an order parameter which has a finite value that continuously vanishes to zero at a critical point. Though the thermodynamic average of the order parameter is zero above the critical point, its fluctuations are non-zero, giving rise to critical phenomena near the phase transition \cite{QPTVojta}. These fluctuations cause a breakdown of the normal macroscopic laws and as a consequence, all observables depend via power laws, which in turn define classical critical exponents, also known as Wilson exponents \cite{mucio,mucio2}. The set of these critical exponents completely characterizes the classical critical behavior near the phase transition, depending only on the dimensionality and symmetry of the order parameter \cite{QPTVojta}. In this way, the microscopic details of a particular phase transition becomes unimportant and all physical systems displaying critical behavior that share the same symmetries and dimensionality, will also share the same critical exponents, called universality class, regardless of the specific magnitude and nature of the microscopic interactions \cite{Scaling,Collins}.

Another kind of second-order phase transitions can occur at $T=0$, where a non-thermal control parameter such as magnetic field, chemical substitution or pressure may drive the system to a critical point, namely a QCP \cite{Sondhi}. Though one cannot access experimentally $T=0$, quantum fluctuations are known to affect many thermodynamic observables well above the QCP. This phenomena has been commonly observed in materials presenting antiferromagnetic \cite{mucio,mucio2,RevLohneysen}, spin density wave \cite{SCQCP} or superconducting \cite{scqcp} phase transitions, in which the ordering temperature is suppressed to zero at a QCP by a non-thermal control parameter. In these systems the presence of quantum fluctuations is evidenced by the collapse of the conventional quasi-particle excitations, giving rise to non-Fermi liquid behavior and unusual power laws of the observables near the quantum critical region \cite{QPTVojta}.

As one approaches a quantum phase transition asymptotically, it is possible to describe the quantum critical behavior classically, since this transition is related to a classical analog in a different spatial dimensionality \cite{QPTVojta,Sondhi}. The effective dimensionality of a quantum phase transition is given by $d_{eff}=(d+z)$, where $d$ is the space dimension and $z$ is a dynamic critical exponent. If $d_{eff}$ is equal to or exceeds the upper critical dimension $d^{+}_{c}=4$ (for most critical phenomena) all the critical exponents reach their classical mean-field values \cite{QPTVojta,mucio,mucio2,Vasin}, for which the order parameter critical exponent is $\beta=0.5$ \cite{Collins}. This is a direct indication of the increase of $d_{eff}$  due to quantum fluctuations, while the system's universality class remains intact \cite{Vasin}. This critical behavior has been suggested to occur in the antiferromagnetic materials MnCl$_2$$\cdot$4H$_2$O \cite{Erkelens} and EuTe \cite{Stishov}.

In this work, we extract the CDW order parameter critical exponent $\beta$ of two cubic superconducting intermetallic CDW systems, (Sr$_{1-x}$Ca$_x$)$_3$Ir$_4$Sn$_{13}$ and Lu(Pt$_{1-x}$Pd$_{x}$)$_2$In as a function of pressure and chemical substitution respectively, by means of x-ray diffraction measurements. There are very few cases of CDW order existing in 3D materials, such as the cubic CuV$_2$S$_4$ \cite{CuV2S4}. Thus, extending the study of the CDW dynamics to other 3D compounds is highly desirable. The CDW order parameter is proportional to the square root of the CDW diffraction peak integrated intensity ($\propto\sqrt{I_{CDW}}$). By mapping the temperature dependence of $I_{CDW}$ we observe that as both systems approach the critical point, where $T_{\rm CDW}\rightarrow0$, $\beta$ continuously changes its value from the expected 3D dimensionality to a mean-field value, supporting the increase of $d_{eff}$ by quantum fluctuations. This provides evidence for the existence of a QCP in these intermetallic CDW materials, rather than relying only on unusual power laws of thermodynamic observables and/or enhancement of superconductivity.

In the (Sr$_{1-x}$Ca$_x$)$_3$Ir$_4$Sn$_{13}$ cubic compounds ($Pm\bar{3}n$ space group), pressure and chemical substitution were combined to suppress the CDW phase and it was argued that a linear temperature dependence of the electrical resistivity near the point where $T_{\rm CDW}$ vanishes would be related with a CDW-QCP, which occurs at a critical pressure of $P_c \sim 18$ kPa \cite{qcp3413}. The same argument was used for (Sr$_{1-x}$Ca$_x$)$_3$Rh$_4$Sn$_{13}$ \cite{qcp34132} where the QCP is believed to be at an optimum substitution of $x_c=0.9$, though more recent lattice dynamics studies also show evidence for quantum criticality \cite{qcp34133}.

For the cubic Heusler Lu(Pt$_{1-x}$Pd$_{x}$)$_2$In ($L2_1$ structure; $Fm\bar{3}m$ space group), the parent compound LuPt$_2$In presents a robust partially gaped CDW transition which sets in at $T_{\rm CDW}=497$ K and a superconducting state at $T_c=0.45$ K \cite{CDWQCP}. Increasing the Pd content also enhances $T_c$ that reaches a sharp maximum at $x_c=0.58$, where simultaneously the CDW state vanishes since $T_{\rm CDW}$ is tuned to zero. Evidences of quantum fluctuations were seen in unusual power laws of the magnetic susceptibility, electrical resistivity and specific heat, as the QCP is approached \cite{CDWQCP}. Additionally, neutron diffraction experiments on the parent compound reveal an order parameter critical exponent of $\beta=0.31\pm0.09$, which is the expected value for a typical three-dimensional universality class \cite{CDWQCP,Collins}.

\section{Experimental details}

Polycrystalline samples of Lu(Pt$_{1-x}$Pd$_{x}$)$_2$In ($x=0.3$, 0.4 and 0.5) were synthesized using an arc-melting furnace under argon atmosphere and pellets were annealed for 150 h at 750$^\circ$C \cite{CDWQCP}. An amount of each sample were grinded and the powder was subsequently annealed for another 100 h at the same temperature before x-ray diffraction characterization. Measurements of electron dispersive spectroscopy were carried out to confirm the compositions (not shown). Flux-grown single crystals of Sr$_3$Ir$_4$Sn$_{13}$ and Ca$_3$Ir$_4$Sn$_{13}$ were synthesized using Sn excess, as reported elsewhere \cite{cecosn}. Phase purity was checked by conventional powder x-ray diffraction (PXRD) on all produced samples (not shown). Temperature dependent four-probe method electrical resistivity were measured in a Quantum Design PPMS DynaCool for primarily sample characterization. Synchrotron PXRD measurements on Lu(Pt$_{1-x}$Pd$_{x}$)$_2$In samples were performed at beamline XPD of the Brazilian Synchrotron Light Source (LNLS). The samples were placed in a cold-finger of a closed-cycle He cryostat. The synchrotron PXRD patterns were obtained by performing $\theta-2\theta$ scans and the intensity was collected using a linear Mythen detector with an angular window of 3.5$^\circ$. The beamline energy ($E$) was set to 10 keV.  High-pressure single crystal x-ray diffraction (SCXRD) experiments on Sr$_3$Ir$_4$Sn$_{13}$ and Ca$_3$Ir$_4$Sn$_{13}$ were performed at beamline XDS ($E=20$ keV) of the LNLS \cite{xds} and beamline P07 ($E=98.7$ keV) of the Petra III, DESY \cite{desy}, respectively. For details of the SCXRD experiments against pressure see Ref. \onlinecite{3413arxiv}.

\section{Results and discussion}

\begin{figure}
\begin{center}
\includegraphics[width=0.5\textwidth]{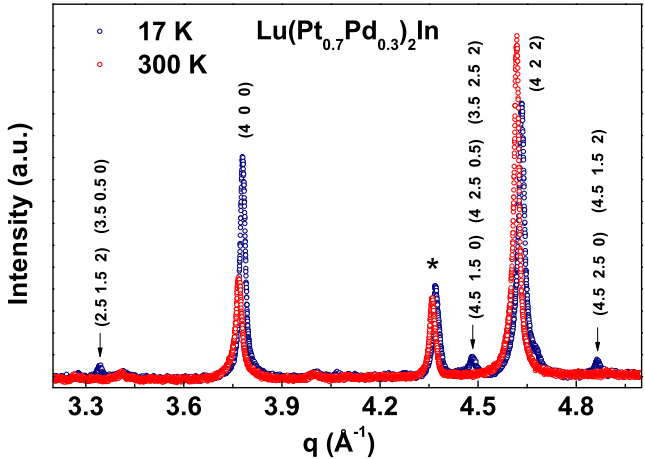}
\end{center}
\caption{Synchrotron PXRD pattern of Lu(Pt$_{1-x}$Pd$_{x}$)$_2$In at $T=17$ and 300 K ($E=10$ keV). CDW/superlattice diffraction peaks are indicated by arrows. A spurious phase Bragg peak marked with $*$ was detected (see main text).}
\label{Fig1}
\end{figure}

In the x-ray diffraction measurements, the CDW/superlattice modulation gives rise to satellites peaks below $T_{\rm CDW}$, which intensities increase as the temperature decreases. For (Sr$_{1-x}$Ca$_x$)$_3$(Ir,Rh)$_4$Sn$_{13}$ and other related compounds, it is well established that the CDW transition doubles the cubic lattice parameter in respect to the higher temperature phase, with a commensurate propagation vector type $\mathbf{q_{\rm CDW}}=(0.5, 0.5, 0)$ \cite{3413arxiv,neutrons,La3Co4Sn13}. By analysing the positions of the superlattice satellite peaks of our PXRD data for Lu(Pt$_{1-x}$Pd$_{x}$)$_2$In (see Fig. \ref{Fig1}), we are able to index these peaks with the same $\mathbf{q_{\rm CDW}}=(0.5, 0.5, 0)$ of (Sr$_{1-x}$Ca$_x$)$_3$(Ir,Rh)$_4$Sn$_{13}$, as also indicated by inelastic neutron scattering experiments \cite{SCES}. The data also suggests that the modulation vector remains constant in the whole inspected temperature range (not shown). A spurious phase Bragg peak was detected, as previously reported in Ref. \onlinecite{CDWQCP}, most likely related to the (220) Bragg reflection of cubic LuX$_{3}$ (X = Pt, Pd, In) ($Pm\bar{3}m$ space group), with no appreciable temperature dependence observed.

\begin{figure}
\begin{center}
\includegraphics[width=0.5\textwidth]{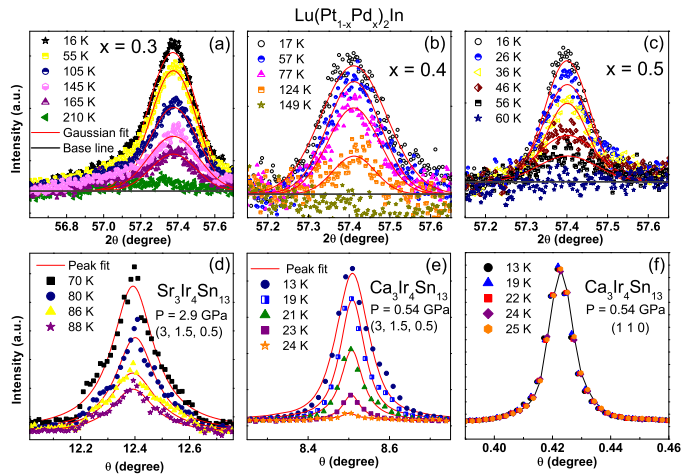}
\end{center}
\caption{Temperature dependence of the superlattice reflection for selected temperatures. Ambient pressure synchrotron PXRD of Lu(Pt$_{1-x}$Pd$_{x}$)$_2$In (4.5, 1.5, 2) reflection for (a) $x=0.3$, (b) $x=0.4$ and (c) $x=0.5$. Solid lines are Gaussian fits. Representative SCXRD of the superlattice peak of (d) Sr$_3$Ir$_4$Sn$_{13}$ (3, 2.5, 0.5) reflection at $P=2.9$ GPa and (e) Ca$_3$Ir$_4$Sn$_{13}$ (3, 1.5, 0.5) reflection at $P=0.54$ GPa. Solid lines are Lorentzian fits. (f) Temperature dependence of the (1, 1, 0) Bragg peak for Ca$_3$Ir$_4$Sn$_{13}$ at $P=0.54$ GPa. Solid line is guide to the eyes.}
\label{Fig2}
\end{figure}

The intensities of the (4.5, 1.5, 2) superlattice peak at $2\theta \approx 57.4^\circ$ ($Q_{\rm CDW} \approx 4.864$ {\AA}$^{-1}$), studied by synchrotron PXRD, for Lu(Pt$_{1-x}$Pd$_{x}$)$_2$In ($x=0.3$, 0.4 and 0.5) are followed as a function of temperature (at ambient pressure) [Figs. \ref{Fig2}(a)-(c)]. There is a clear temperature dependence of the intensity of the superlattice peaks, which disappears at around $T_{\rm CDW}$. The same behavior, measured by SCXRD, is seen for the superlattice peaks (3, 2.5, 0.5) of Sr$_3$Ir$_4$Sn$_{13}$ and (3, 1.5, 0.5) of Ca$_3$Ir$_4$Sn$_{13}$, at $P=2.9$ [Fig. \ref{Fig2}(d)] and  $P=0.54$ GPa [Fig. \ref{Fig2}(e)], respectively.  Figs. \ref{Fig2}(d)-(f) are representative of the behavior at all pressure range measured for the (Sr,Ca)$_3$Ir$_4$Sn$_{13}$ materials. As a comparison, we show that the temperature dependence of the (1, 1, 0) Bragg reflection for Ca$_3$Ir$_4$Sn$_{13}$ at $P=0.54$ GPa [Fig. \ref{Fig2}(f)], where no perceivable changes occur as the compound goes through the CDW transition.

\begin{figure}
\begin{center}
\includegraphics[width=0.5\textwidth]{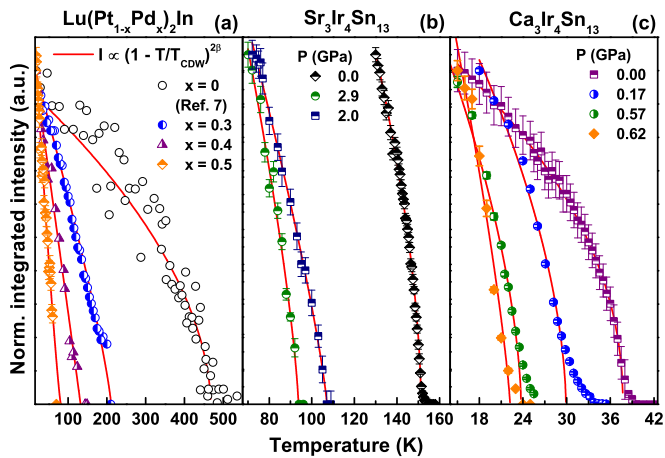}
\end{center}
\caption{Temperature dependence of the integrated intensity of the superlattice reflection, normalized at $T\sim15$ K. The temperature dependence was fitted to the power law $I\propto{(1-T/T_{\rm CDW})}^{2\beta}$ (red solid lines). (a) Lu(Pt$_{1-x}$Pd$_{x}$)$_2$In. Data for $x=0$ is from Ref. \onlinecite{CDWQCP}. (b) Sr$_3$Ir$_4$Sn$_{13}$. Data from Ref. \onlinecite{3413arxiv}. (c) Ca$_3$Ir$_4$Sn$_{13}$. Data from Ref. \onlinecite{3413arxiv}}
\label{Fig3}
\end{figure}

The onset of the CDW order parameter can be directly tracked by measuring the evolution of the integrated intensity of the superlattice reflection as a function of temperature. In this way, one may use the power law $I_{CDW}\propto{(1-T/T_{\rm CDW})}^{2\beta}$, fitted around $T_{\rm CDW}$, to obtain the CDW order parameter critical exponent $\beta$. This analysis is depicted in Fig. \ref{Fig3} for Lu(Pt$_{1-x}$Pd$_{x}$)$_2$In as a function of chemical substitution [Fig. \ref{Fig3}(a)], and for Sr$_3$Ir$_4$Sn$_{13}$ [Fig. \ref{Fig3}(b)] and Ca$_3$Ir$_4$Sn$_{13}$ [Fig. \ref{Fig3}(c)] against pressure. One can observe that in Sr$_3$Ir$_4$Sn$_{13}$ the curve shape does not change drastically by increasing pressure. In contrast, for Lu(Pt$_{1-x}$Pd$_{x}$)$_2$In and Ca$_3$Ir$_4$Sn$_{13}$ a clear modification of the temperature dependence of the integrated intensity occurs as a function of chemical doping and pressure, respectively, thus affecting $\beta$ as it will be discussed later. These results were obtained on warming the samples through the CDW/superlattice transition. We have, in addition, performed measurements on cooling for all samples and no appreciable thermal hysteresis, within the instrument resolution, was observed, implying that the CDW phase transition remains second-order as $T_{\rm CDW}\rightarrow0$.

\begin{figure}
\begin{center}
\includegraphics[width=0.5\textwidth]{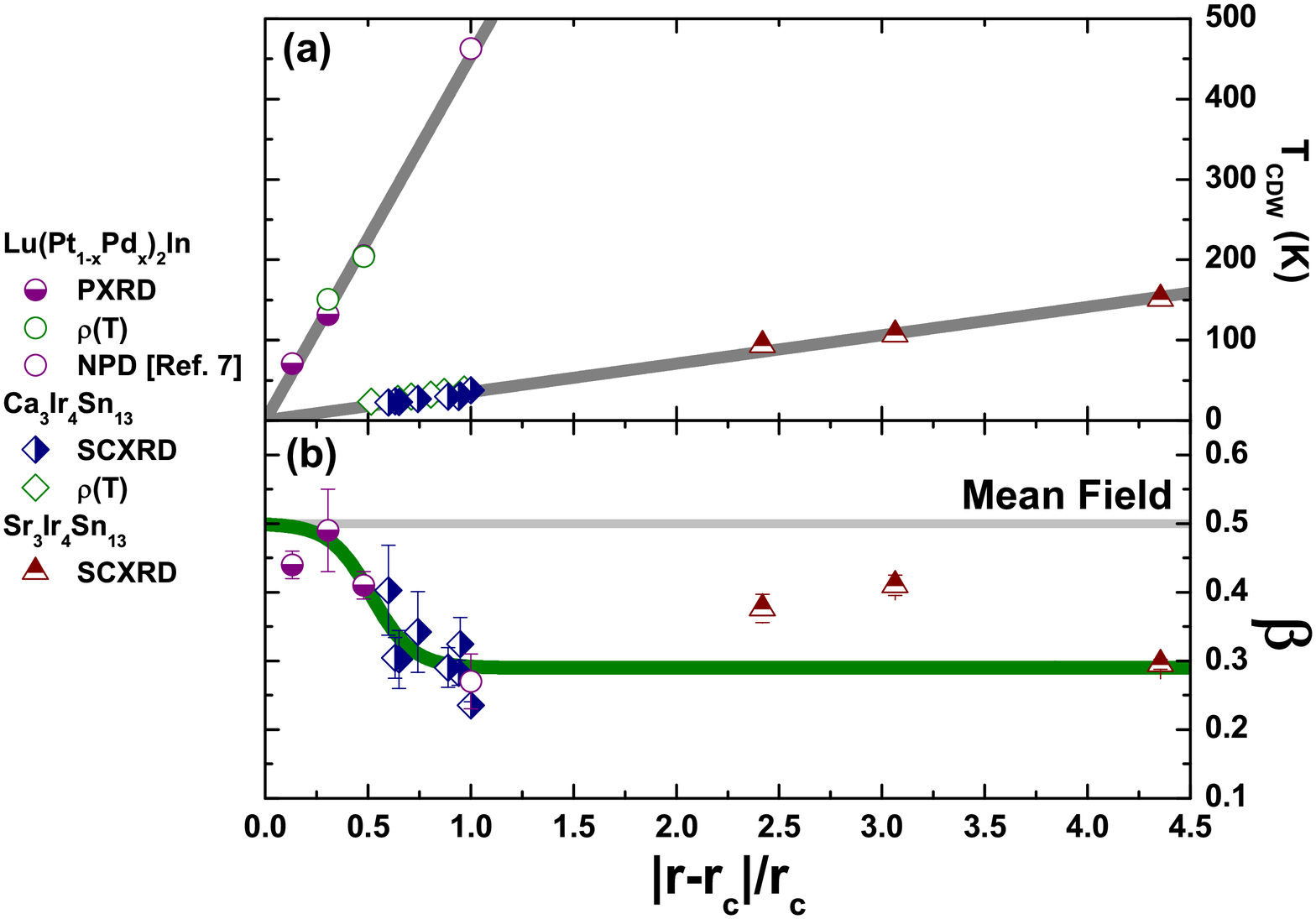}
\end{center}
\caption{Phase diagram of the Lu(Pt$_{1-x}$Pd$_{x}$)$_2$In, Sr$_3$Ir$_4$Sn$_{13}$ and Ca$_3$Ir$_4$Sn$_{13}$ systems as a function of the control parameter $r$, normalized by the critical control parameter $r_c$ (see main text). (a) $T_{\rm CDW}$ against the control parameter, obtained by x-ray diffraction and electrical resistivity (not shown). Solid lines are linear fits. Data for LuPt$_{2}$In is from Ref. \onlinecite{CDWQCP}, obtained by neutron powder diffraction (NPD). (b) CDW order parameter critical exponent $\beta$ versus the control parameter. Solid line is guide to the eyes.}
\label{Fig4}
\end{figure}

With $T_{\rm CDW}$ and CDW order parameter critical exponent extracted from the analysis presented in Fig. \ref{Fig3}, we construct the phase diagram shown in Fig. \ref{Fig4}(a). Here we present these parameters on both Lu(Pt$_{1-x}$Pd$_{x}$)$_2$In and (Sr,Ca)$_3$Ir$_4$Sn$_{13}$ cubic compounds. While $T_{\rm CDW}$ for Lu(Pt$_{1-x}$Pd$_{x}$)$_2$In is suppressed to zero by chemical substitution, the non-thermal control parameter ($r$) in (Sr,Ca)$_3$Ir$_4$Sn$_{13}$ is pressure. In this sense, we have normalized $r$ of both systems by its value when $T_{\rm CDW}=0$,  which were extracted from the linear fittings of the phase diagram in Fig. \ref{Fig4}(a). The quantum critical points were estimated as being, respectively, $x_c\sim0.61$ (Pd content) and $P_c\sim 1.55$ GPa. At ambient pressure, Sr$_3$Ir$_4$Sn$_{13}$ represents a negative pressure of $P=-5.2$ GPa in respect to $P_c$ of Ca$_3$Ir$_4$Sn$_{13}$ \cite{qcp3413}. Thus, one can adopt the $|r-r_c|/r_c$ ratio to measure the distance to $T_{\rm CDW}=0$ for both systems. In Fig. \ref{Fig4} the QCP is represented when $|r-r_c|/r_c=0$. Values of $|r-r_c|/r_c>1$ indicate that the system is far from the QCP, meaning that within the studied pressure range, Sr$_3$Ir$_4$Sn$_{13}$ was always in the negative pressure limit. Fig. \ref{Fig4}(b) shows that the CDW order parameter critical exponent for Sr$_3$Ir$_4$Sn$_{13}$ stays well inside the three-dimensional characteristic value. On the other hand, for Lu(Pt$_{1-x}$Pd$_{x}$)$_2$In and Ca$_3$Ir$_4$Sn$_{13}$, $T_{\rm CDW}$ could be driven closer to zero. In these latter systems there is a striking difference of the order parameter critical exponent dependence on $r$, in which both start with $\beta\sim0.3$, characteristic of three-dimensional universality classes, and monotonically increases its value towards $\beta\sim0.5$.

The monotonic behavior, as shown by Vasin \textit{et al.} \cite{Vasin}, is caused by the fact that the effective dimension has a dependence of the form $d_{eff}=d+z\Lambda(\omega/k_BT)$, where $\Lambda(\omega/k_BT)$ is a continuous smooth function of the ratio between the classical and quantum scales (see Ref. \onlinecite{Vasin}). In the limit where the frequency of quantum fluctuations is much lower than thermal fluctuations ($\omega<<k_BT$), it corresponds to the classical thermal-driven second-order phase transitions. On the other hand, when $\omega>>k_BT$ the critical fluctuations has a quantum nature.

The question whether the suppression of $T_{\rm CDW}$ in Lu(Pt$_{1-x}$Pd$_{x}$)$_2$In by chemical doping and in (Sr$_{1-x}$Ca$_x$)$_3$Ir$_4$Sn$_{13}$ by pressure is in fact driving these systems to a QCP can now be addressed. The change of the order parameter critical exponent from $\beta\sim0.3$ to $\beta\sim0.5$ is an indication of the effective dimensionality increase due to quantum critical fluctuations by adding a dynamic exponent and attaining the condition where $d_{eff}=(d+z)\geq d^{+}_{c}$, in which critical exponents reach its mean-field values \cite{QPTVojta,mucio,mucio2,Collins}. This is a direct evidence of the presence of a quantum phase transition at the QCP ($|r-r_c|/r_c=0$), where $T_{\rm CDW}=0$. While we were not able to follow $T_{\rm CDW}$ down to very low temperatures, the increase in the effective dimensionality promotes a classical-to-quantum crossover in the vicinity of the QCP that already sets in at appreciable temperatures \cite{Varma}. One may argue that disorder may also affect the critical exponents, though this would imply in a smeared out QCP, which is not observed for both systems [Fig. \ref{Fig4}(a)] \cite{qcp3413,CDWQCP}.

By mapping the CDW order parameter critical exponent through a quantum phase transition, one has a more reliable thermodynamic parameter to attest the dominance of quantum fluctuations. It is yet to be studied the role of these quantum fluctuations on the superconducting state in these compounds and whether indeed the presence of a QCP enhances $T_c$. Nevertheless, we clearly show compelling evidence that a quantum phase transition is present in Lu(Pt$_{1-x}$Pd$_{x}$)$_2$In and (Sr$_{1-x}$Ca$_x$)$_3$Ir$_4$Sn$_{13}$ systems and offer an opportunity to explore QCP resulting solely from a tunable, non-magnetic, structural phase transition.

\section{Summary}

In summary, we have studied two cubic superconducting intermetallic systems Lu(Pt$_{1-x}$Pd$_{x}$)$_2$In and (Sr$_{1-x}$Ca$_x$)$_3$Ir$_4$Sn$_{13}$, which present a vanishing CDW transition, under chemical doping and pressure, respectively. For both systems we were able to determine the presence of quantum fluctuations due to a QCP at $T_{\rm CDW}=0$, by following the CDW order parameter critical exponent as function of the control parameter. This approach can be used to explore other CDW/superconducting compounds, such as (Sr$_{1-x}$Ca$_x$)$_3$Rh$_4$Sn$_{13}$ \cite{qcp34132} and R$_2$T$_3$X$_5$ (R = rare-earth elements; T = transition metal, and X = $s$-$p$ metal) \cite{Lu}, and test whether the suppression of $T_{\rm CDW}$ leads to a quantum phase transition. The experimental study of quantum criticality would be strengthened by probing directly the order parameter to gain insights on the role of quantum fluctuations in the physical observables/phenomena rather than relying only on unusual power laws.

\begin{acknowledgments}
We thank E. C. Andrade, M. A. Continentino, F. A. Garcia and T. Micklitz for critical reading of the manuscript and valuable suggestions. This work was supported by the Brazilian funding agencies: Funda\c{c}\~{a}o Carlos Chagas Filho de Amparo \`{a} Pesquisa do Estado do Rio de Janeiro (FAPERJ) [Grant No. E-26/202.798/2019] and Conselho Nacional de Desenvlovimento Cient\'{\i}fico e Tecnol\'{o}gico (CNPq) [Grant No. 400633/2016-7]. L.S.I. Veiga is supported by the by the UK Engineering and Physical Sciences Research Council (Grants No. EP/N027671/1 and No. EP/N034694/1). The authors thank the beamlines XPD and XDS staff for technical support and LNLS for the for concession of beam time [Proposals Nos. 20160202, 20180225 and 20190008]. We also express our gratitude to O. Gutowski and M. v, Zimmermann for their assistance at beamline P07 of Petra III. Part of this research was carried out at PETRA III synchrotron radiation source at DESY, a member of Helmholtz Association (HGF).
\end{acknowledgments}


\begin{thebibliography}{99}
\bibitem{Cu1} J. Chang, E. Blackburn, A. T. Holmes, N. B. Christensen, J. Larsen, J. Mesot, Ruixing Liang, D. A. Bonn, W. N. Hardy, A. Watenphul, M. v. Zimmermann, E. M. Forgan, and S. M. Hayden, Nat. Phys. \textbf{8}, 871 (2012).
\bibitem{Cu2} E. Fradkin, S. A. Kivelson, and J. M. Tranquada, Rev. Mod. Phys. \textbf{87}, 457 (2015).
\bibitem{Cu3} B. Loret, N. Auvray, Y. Gallais, M. Cazayous, A. Forget, D. Colson, M.-H. Julien, I. Paul, M. Civelli, and A. Sacuto, Nat. Phys. \textbf{15}, 771 (2019).
\bibitem{Cu4} H. Miao, G. Fabbris, C. S. Nelson, R. Acevedo-Esteves, Y. Li, G. D. Gu, T. Yilmaz, K. Kaznatcheev, E. Vescovo, M. Oda, K. Kurosawa, N. Momono, T. A. Assefa, I. K. Robinson, J. M. Tranquada, P. D. Johnson, and M. P. M. Dean, arXiv:2001.10294.
\bibitem{qcp3413} L. E. Klintberg, S. K. Goh, P. L. Alireza, P. J. Saines, D. A. Tompsett, P. W. Logg, J. Yang, B. Chen, K. Yoshimura, and F. M. Grosche, Phys. Rev. Lett. \textbf{109}, 237008 (2012).
\bibitem{qcp34132} S. K. Goh, D. A. Tompsett, P. J. Saines, H. C. Chang, T. Matsumoto, M. Imai, K. Yoshimura, and F. M. Grosche, Phys. Rev. Lett. \textbf{114}, 097002 (2015).
\bibitem{CDWQCP} T. Gruner, D. Jang, Z. Huesges, R. Cardoso-Gil, G. H. Fecher, M. M. Koza, O. Stockert, A. P. Mackenzie, M. Brando, and C. Geibel, Nat. Phys. \textbf{13}, 967 (2017).
\bibitem{2H-NbSe} Y. Feng, J. Wang, R. Jaramillo, J. v. Wezel, S. Haravifard, G. Srajer, Y. Liu, Z.-A. Xu, P. B. Littlewood, and T. F. Rosenbaum, Proc. Natl. Acad. Sci. USA \textbf{109}, 7224 (2012).
\bibitem{Lu} N. S. Sangeetha, A. Thamizhavel, C. V. Tomy, Saurabh Basu, A. M. Awasthi, S. Ramakrishnan, and D. Pal, Phys. Rev. B \textbf{86}, 024524 (2012).
\bibitem{QPTVojta} M. Vojta,  Rep. Prog. Phys. \textbf{66}, 2069 (2003).
\bibitem{mucio} M. A. Continentino, Braz. J. Phys. \textbf{35}, 197 (2005).
\bibitem{mucio2} M. A. Continentino, \textit{Quantum scaling in many-body systems} (Cambridge University Press, Cambridge, 2017).
\bibitem{Scaling} H. E. Stanley, Rev. Mod. Phys. \textbf{71}, S358 (1999).
\bibitem{Collins} M. F. Collins, \textit{Magnetic critical scattering} (Oxford University Press, New York, 1989).
\bibitem{Sondhi} S. L. Sondhi, S. M. Girvin, J. P. Carini, and D. Shahar, Rev. Mod. Phys. \textbf{69}, 315 (1997).
\bibitem{RevLohneysen} H. v. L\"{o}hneysen, A. Rosch, M. Vojta, and P. W\"{o}lfle, Rev. Mod. Phys. \textbf{79}, 1015 (2007).
\bibitem{SCQCP} L. Taillefer, Annu. Rev. Condens. Matter Phys. \textbf{1}, 51 (2010).
\bibitem{scqcp} S. M. Ramos, M. B. Fontes, E. N. Hering, M. A. Continentino, E. Baggio-Saitovich, F. D. Neto, E. M. Bittar, P. G. Pagliuso, E. D. Bauer, J. L. Sarrao, and J. D. Thompson, Phys. Rev. Lett. \textbf{105}, 126401 (2010).
\bibitem{Vasin} M. Vasin, V. Ryzhov and V. M. Vinokur, Sci. Rep. \textbf{5}, 18600 (2015).
\bibitem{Erkelens} W. A. C. Erkelens, L. P. Regnault, J. Rossat-Mignod, J. E. Moore, R. A. Butera, and L. J. de Jongh, Europhys. Lett. \textbf{l}, 37 (1986).
\bibitem{Stishov} S. M. Stishov, A. E. Petrova, S. Yu. Gavrilkin, and L. A. Klinkova, Phys. Rev. B \textbf{91}, 144416 (2015).
\bibitem{CuV2S4} R. M. Fleming, F. J. DiSalvo, R. J. Cava, and J. V. Waszczak, Phys. Rev. B \textbf{24}, 2850 (1981).
\bibitem{qcp34133} Y. W. Cheung, Y. J. Hu, M. Imai, Y. Tanioku, H. Kanagawa, J. Murakawa, K. Moriyama, W. Zhang, K. T. Lai, K. Yoshimura, F. M. Grosche, K. Kaneko, S. Tsutsui, and S. K. Goh, Phys. Rev. B \textbf{98}, 161103(R) (2018).
\bibitem{cecosn}  C. Israel, E. M. Bittar, O. E. Ag\"{u}ero, R. R. Urbano, C. Rettori, I. Torriani, P. G. Pagliuso, N. O. Moreno, J. D. Thompson, M. F.
Hundley, J. L. Sarrao, and H. A. Borges, Physica B \textbf{359-361}, 251 (2005).
\bibitem{xds} F. A. Lima, M. E. Saleta, R. J. S. Pagliuca, M. A. Eleot\'{e}rio, R. D. Reis, J. Fonseca J\'{u}nior, B. Meyer, E. M. Bittar, N. M. Souza-Neto, and E. Granado, J. Synchrotron Rad. \textbf{23}, 1538 (2016).
\bibitem{desy} J. Strempfer, S. Francoual, D. Reuther, D. K. Shukla, A. Skaugen, H. Schulte-Schrepping, T. Kracht, and H. Franz, J. Synchrotron Radiat. \textbf{20}, 541 (2013).
\bibitem{3413arxiv} L. S. I. Veiga, J. R. L. Mardegan, M. v. Zimmermann, D. T. Maimone, F. B. Carneiro, M. B. Fontes, J. Strempfer, E. Granado, P. G. Pagliuso, and E. M. Bittar, Phys. Rev. B \textbf{101}, 104511 (2020).
\bibitem{neutrons} D. G. Mazzone, S. Gerber, J. L. Gavilano, R. Sibille, M. Medarde, B. Delley, M. Ramakrishnan, M. Neugebauer, L. P. Regnault, D. Chernyshov, A. Piovano, T. M. Fern\'{a}ndez-D\'{\i}az, L. Keller, A. Cervellino, E. Pomjakushina, K. Conder, and M. Kenzelmann, Phys. Rev. B \textbf{92}, 024101 (2015).
\bibitem{La3Co4Sn13} L. Mendon\c{c}a-Ferreira, F. B. Carneiro, M. B. Fontes, E. Baggio-Saitovitch, L. S. I. Veiga, J. R. L. Mardegan, J. Strempfer, M. M. Piva, P. G. Pagliuso, R. D. dos Reis, and E. M. Bittar, J. Alloys Compds. \textbf{773}, 34 (2019).
\bibitem{SCES} T. Gruner, S. Lucas, S. Tsutsui, K. Kaneko, K. Schmalzl, M. M. Koza, C. Geibel, and O. Stockert, \textit{Critical Phonon Softening Near a Structural Instability in the Quantum Critical System Lu(Pt$_{1-x}$Pd$_{x}$)$_2$In} (International Conference on Strongly Correlated Electron Systems, Okayama, 2019).
\bibitem{Varma} C. M. Varma, Z. Nuzzinov, W. V. Saarloos, Phys. Rep. \textbf{361}, 267 (2002).
\end{thebibliography}
\end{document}